\documentclass[fleqn,12pt]{article}
\usepackage{epsfig}
\newcommand{\be}{\begin{equation}}
\newcommand{\ee}{\end{equation}}

\newcommand{\eq}[1]{eq.~(\ref{#1})}

\newcommand{\nc}{\newcommand}
\nc{\rnc}{\renewcommand}
\nc{\bc}{\begin{center}}
\nc{\ec}{\end{center}}
\nc{\bfl}{\begin{flushleft}}
\nc{\efl}{\end{flushleft}}
\nc{\bfr}{\begin{flushright}}
\nc{\efr}{\end{flushright}}
\nc{\bi}{\begin{itemize}}
\nc{\ei}{\end{itemize}}
\nc{\ben}{\begin{enumerate}}
\nc{\een}{\end{enumerate}}
\nc{\bd}{\begin{description}}
\nc{\ed}{\end{description}}

\newfont{\cmmibxii}{cmmib10 scaled \magstep 1}
\newfont{\cmmibx}{cmmib10}
\newfont{\cmmibix}{cmmib9}
\newcommand{\fig}[1]{Fig.\,\ref{#1}}

\newcommand{\pbar}{\bar p}

\newcommand{\alphabold}{\mbox{\small\boldmath $\alpha$}}
\newcommand{\xbold}{\mbox{\boldmath $x$}}

\newcommand{\delchisq}{\Delta \chi^2_i(x_i;\alphabold)}
\newcommand{\delchi}{\Delta \chi^2_i}

\newcommand{\delchimax}{{\delchi}_{\rm max}}
\newcommand{\x}{(\nu/m)}
\newcommand{\y}{(\nu_0/m)}

\def\lpa{\lambda_{p{-}\rm air}}
\def\spa{\sigma_{p{-}\rm air}}
\def\spai{\sigma_{p{-}\rm air}^{\rm inel}}
\def\spae{\sigma_{p{-}\rm air}^{\rm el}}
\def\spaqe{\sigma_{p{-}\rm air}^{q{-}\rm el}}
\begin{document}    
\renewcommand\thepage{\ }
%
%
\begin{titlepage} 
%
\newcommand\reportnumber{1105} 
\newcommand\mydate{June 1, 2006} 
\newlength{\nulogo} 
\settowidth{\nulogo}{\small\sf{N.U.H.E.P. Report No. \reportnumber}}
\title{\hfill\fbox{{\parbox{\nulogo}{\small\sf{Northwestern University: \\
N.U.H.E.P. Report No. \reportnumber\\ \mydate
}}}}\vspace{1in} \\
{The Elusive p-air Cross Section}}
\author{Martin M. Block
 \\
{\small\em Department of Physics and Astronomy,} \vspace{-5pt} \\ 
{\small\em Northwestern University, Evanston, IL 60208}
\vspace{-5pt}\\
{\small\em e-mail:}{\footnotesize \sf \ mblock@northwestern.edu}\vspace{5pt}\\
{\footnotesize \sf Paper delivered at c2cr2005 Conference, Prague,  September 7-13, 2005}
\vspace{.05in}\\
}
\vfill
\vspace{.5in}
\date{} 
\maketitle
\begin{abstract}
For the $\pbar p$ and $pp$ systems, we have used all of the extensive data of the Particle Data Group[K. Hagiwara {\em et al.} (Particle Data Group), Phys. Rev. D 66, 010001 (2002).]. We then subject these data to a screening process, the ``Sieve'' algorithm[M. M. Block, physics/0506010.], in order to eliminate ``outliers'' that can skew a $\chi^2$ fit. With the ``Sieve'' algorithm, a robust fit using a Lorentzian distribution is first made to all of the data to sieve out abnormally high $\delchi$, the individual i$^{\rm th}$ point's contribution to the total $\chi^2$. The $\chi^2$ fits are then made to the sieved data.
We demonstrate that we cleanly discriminate between asymptotic $\ln s$ and $\ln^2s$ behavior of total hadronic cross sections when we require that these amplitudes {\em also} describe, on average, low energy data dominated by resonances. We simultaneously fit real analytic amplitudes to the ``sieved'' high energy measurements of  $\bar p p$ and $pp$ total cross sections and $\rho$-values  for $\sqrt s\ge 6$ GeV, while requiring that their asymptotic fits smoothly  join  the   the $\sigma_{\bar p p}$ and $\sigma_{pp}$ total  cross sections at $\sqrt s=$4.0 GeV---again {\em both} in magnitude and slope.  Our results strongly favor  a high energy $\ln^2s$ fit, basically excluding a $\ln s$ fit. Finally, we make a screened Glauber fit for the p-air cross section, using as input our precisely-determined $pp$ cross sections at cosmic ray energies.  
\end{abstract}

\end{titlepage} 
%
\pagenumbering{arabic}
\renewcommand{\thepage}{-- \arabic{page}\ --}  
%
\section{Introduction}
This paper consists of 3 parts:
\begin{enumerate}
\item The introduction of the ``Sieve'' algorithm of M. M. Block\cite{sieve}, used to ``sieve'' out ``outliers'' from the Particle Data Group\cite{pdg} compilations of cross sections and $\rho-$values (the ratio of the real to the imaginary portion of the forward scattering amplitude).
\item A new, constrained  method\cite{newFroissart} of  fitting $\bar p$ and $pp$ cross sections and $\rho-$values, showing that the Froissart bound is saturated. This work was done in conjunction with Francis Halzen.
\item A Glauber calculation for the p-air production cross section, using screening, an unpublished work done in collaboration with Ralph Engel.  
\end{enumerate}
\section{Part 1: The ``Sieve'' Algorithm}
We now outline the adaptive Sieve  algorithm\cite{sieve} that minimizes the effect that ``outliers''---points with abnormally high contributions to $\chi^2$---have on a fit when they contaminate a data sample that is otherwise Gaussianly distributed.  Our fitting procedure consists of several steps:
\begin{enumerate}
\item{Make a robust fit  of {\em all} of the data (presumed outliers and all)\ by minimizing $\Lambda^2_0$, the Lorentzian squared with respect to $\alphabold$, where 
\be
\Lambda^2_0(\alphabold;\xbold)\equiv\sum_{i=1}^N\ln\left\{1+0.179\delchisq\right\},\label{lambda0}
\ee 
 with  $\alphabold=\{\alpha_1,\ldots,\alpha_M\}$ being the $M$-dimensional parameter space of the fit. $\xbold=\{{x_1,\ldots,x_N}\}$ represents the abscissa of the $N$ experimental measurements $\mbox{\boldmath $y$}=\{y_1,\ldots,y_N\}$ that are  being fit and $\delchisq\equiv \left(\frac{y_i-y(x_i;\alphabold)}{\sigma_i}\right)^2$ is the individual $\chi^2$ contribution of the $i^{\rm th}$ point,  where $y(x_i;\alphabold)$ is the theoretical value at $x_i$ and $\sigma_i$ is the experimental error. It is shown in ref. \cite{sieve} that for Gaussianly distributed data, minimizing $\Lambda^2_0$ gives, on average,  the same total $\chi^2_{\rm min}\equiv\sum_{i=1}^N \delchisq$ from \eq{lambda0} as that found in a conventional $\chi^2$ fit,  as well as  rms widths (errors) for the parameters that are almost the same as those found in a $\chi^2$ fit}. 

A quantitative measure of whether point $i$ is an  outlier, {\em i.e.,} whether it is ``far away'' from the true signal,  is the magnitude of its $\delchisq= \left(\frac{y_i-y(x_i;\alphabold)}{\sigma_i}\right)^2$. The reason for minimizing the Lorentzian squared is that this procedure gives the outliers much less weight $w$ in the fit ($w\propto$ 1/$\sqrt{\delchisq}$\,), for large $\delchisq$) than does a $\chi^2$ fit ($w\propto \sqrt{\delchisq}$\,), thus making the fitted parameters insensitive to  outliers and hence robust. For details, see ref. \cite{sieve}.

If $\chi^2_{\rm min}$ is satisfactory, make a conventional $\chi^2$ fit to get the errors and you are finished.   If $\chi^2_{\rm min}$ is not satisfactory, proceed to step 
 \ref{nextstep}.
\item {Using the above robust $\Lambda^2_0$ fit as the initial estimator for the theoretical curve, evaluate $\delchisq$, for each of the $N$ experimental points.}\label{nextstep}
\item A largest cut, $\delchisq_{\rm max}$, must now be selected. We start the process with $\delchisq_{\rm max}=9$. If any of the points have $\Delta \chi^2_i(x_i;\alphabold)>\delchisq_{\rm max}$, reject them---they fell through the ``Sieve''. The choice of $\delchisq_{\rm max}$ is an attempt to pick  the largest ``Sieve'' size (largest $\delchisq_{\rm max}$) that rejects all of the outliers, while minimizing the number of signal points  rejected. \label{redo}
\item Next, make a conventional $\chi^2$ fit to the sifted set---these data points are the ones that have been retained in the ``Sieve''. This  fit is used to estimate   $\chi^2_{\rm min}$.    Since the data set has been truncated by eliminating the points with $\delchisq>\delchisq_{\rm max}$, we must slightly renormalize the $\chi^2_{\rm min}$ found to take this into account, by the factor $\cal R$.  For $\delchimax=9,6,$ and 4, the factor $\cal R$ is given by 1.027, 1.140 and 1.291, whereas the fraction of the points that should survive this $\chi^2$ cut---for a Gaussian distribution---is 0.9973, 0.9857 and 0.9545, respectively. A plot of ${\cal R}^{-1}$ as a function of $\delchimax$ is given in Figure \ref{renorm}, which is taken from ref. \cite{sieve}.

If the renormalized $\chi^2_{\rm min}$, {\em i.e.,} ${\cal R}\times \chi^2_{\rm min}$ is acceptable---in the {\em conventional} sense, using the ordinary $\chi^2$ distribution probability function---we consider the fit of the data to the  model to be satisfactory  and proceed to the next step. If the renormalized $\chi^2_{\rm min}$ is not acceptable and $\delchisq_{\rm max}$ is not too small, we pick a smaller $\delchisq_{\rm max}$ and go back to step \ref{redo}. The smallest value of $\delchisq_{\rm max}$ that we used is $\delchisq_{\rm max}=4$.  

\item
From the  $\chi^2$ fit that was made to the ``sifted'' data in the preceding step, evaluate  the parameters $\alphabold$.
Next, evaluate the $M\times M$ covariance (squared error) matrix of the parameter space which was found in the $\chi^2$ fit. We find the new squared error matrix for the $\Lambda^2$  fit by multiplying the covariance matrix by the square of the factor $r_{\chi^2}$. From Figure \ref{renorm}, we find that  $r_{\chi^2}\sim 1.02,1.05$ and  1.11  for $\delchisq_{\rm max}=9$, 6 and 4,  respectively . The values of $r_{\chi^2}>1$ reflect the fact that a $\chi^2$ fit to the {\em truncated} Gaussian distribution that we obtain---after first making  a robust fit---has a rms (root mean square) width which is somewhat greater than the  rms width of the $\chi^2$ fit to the same untruncated distribution\cite{sieve}. 
\end{enumerate}

The application of a $\chi^2$ fit to the {\em sifted set} gives stable estimates of the model parameters $\alphabold$, as well as a goodness-of-fit of the data to the model when $\chi^2_{\rm min}$ is renormalized for the effect of truncation due to the cut $\delchisq_{\rm max}.$  One can now use conventional probabilities for $\chi^2$ fits, {\em i.e.,} the probability that $\chi^2$ is greater than ${\cal R}\times\chi^2_{\rm min}$, for the number of degrees of freedom $\nu$. Model parameter errors are found by multiplying the covariance (squared error) matrix of the conventional $\chi^2$ fit by the appropriate factor $(r_{\chi^2})^2$ for the cut $\delchisq_{\rm max}$.

\section{Part 2: Saturating the Froissart Bound} 
High energy cross sections for the scattering of hadrons should be bounded   by $\sigma \sim \ln^2s$,
where $s$ is the square of the cms energy.   This fundamental result is derived from unitarity and analyticity by Froissart\cite{froissart}, who states: ``At forward or backward angles, the modulus of the amplitude behaves at most like $s\ln^2s$, as $s$ goes to infinity.  We can use the optical theorem to derive that the total cross sections behave at most like $\ln^2s$, as $s$ goes to infinity".  In this context, saturating the Froissart bound refers to an energy dependence of the total  cross section rising no more rapidly than  $\ln^2s$.

The question as to whether any of the present day  high energy data for  $\bar pp$ and  $pp$  cross sections saturate the Froissart bound has not been settled; one can not unambiguously  discriminate between asymptotic fits of $\ln s$ and $\ln^2 s$  using high energy data only\cite{bkw,cudell}.  We here point out that this ambiguity is resolved by requiring that the fits to the high energy data smoothly join the cross section and energy dependence obtained by averaging the resonances at low energy. Imposing this duality\cite{igi} condition, we show that only fits to the high energy data behaving as $\ln^2 s$ that smoothly join (in {\em both} magnitude and first derivative) to the low energy 
data at the ``transition energy" (defined as the energy region just after the resonance regions end)  can  adequately describe the highest energy points.  This technique has recently been successfully used by Block and Halzen\cite{BH} to show that the Froissart bound is saturated for the  $\gamma p$ system.

We will use real analytic amplitudes to describe the data. The total cross sections $\sigma$ are found from the optical theorem and $\rho$ is the ratio of the real to the imaginary portion of the forward scattering amplitude. As shown in ref. {\cite{newFroissart}, in the high energy limit where $s\rightarrow2m\nu$,  we can write
$\sigma^{\pm}$ and  $\rho^{\pm}$, along with the cross section derivatives $\frac{d\sigma^{\pm}}{d\x}$, as sums and differences of even and odd amplitudes, {\em i.e.,}
\begin{eqnarray}
\sigma^\pm&{\!\!\! =\!\!\! }&c_0+c_1\ln\left(\frac{\nu}{m}\right)+c_2\ln^2\left(\frac{\nu}{m}\right)+\beta_{\cal P'}\left(\frac{\nu}{m}\right)^{\mu -1}\pm\  \delta\left({\nu\over m}\right)^{\alpha -1},\label{sigmapm}\\
\rho^\pm&{\!\!\! =\!\!\! }&{1\over\sigma^\pm}\left\{\frac{\pi}{2}c_1+c_2\pi \ln\left(\frac{\nu}{m}\right)-\beta_{\cal P'}\cot\left({\pi\mu\over 2}\right)\left(\frac{\nu}{m}\right)^{\mu -1}+\frac{4\pi}{\nu}f_+(0)\right.\nonumber\\
&&\ \ \ \ \ \ \ \ \ \ \ \ \ \ \ \ \ \ \ \ \  
\left.\pm \delta\tan\left({\pi\alpha\over 2}\right)\left({\nu\over m}\right)^{\alpha -1} \right\}\!\!,\label{rhopm}\\
\frac{d\sigma^{\pm}}{d(\x}&{\!\!\! =\!\!\! }&c_1\left\{\frac{1}{\x}\right\} +c_2\left\{ \frac{2\ln\x}{\x}\right\}+\beta_{\cal P'}\left\{(\mu-1)\x^{\mu-2}\right\}  \nonumber\\
&&\ \ \ \ \ \ \ \ \ \ \ \ \ \ \ \ \ \ \ \ \  
\pm \ \delta\left\{(\alpha -1)(\x)^{\alpha - 2}\right\}\label{derivpm}\!,
\end{eqnarray}
where the upper sign is for $pp$ and the lower sign is for $\bar p p$ scattering.
The exponents $\mu$ and $\alpha$ are real. The real constant $f_+(0)$, appearing only in the $\rho$-value,  is the subtraction constant at $\nu=0$ needed to be introduced into a singly-subtracted dispersion relation\cite{bc},\cite{gilman}. We note that  \eq{sigmapm} is linear  in the real  coefficients $c_0, c_1, c_2, \beta_{\cal P'}$ and $\delta$, convenient for a $\chi^2$ fit to the experimental total cross sections and $\rho$-values.  Throughout we will use units of $\nu$ and $m$ in GeV and cross section in mb, where $m$ is the proton mass. 

It is convenient to define, at the transition energy $\nu_0$,  
\begin{eqnarray}
\sigma_{\rm av}&=&\frac{\sigma^{+}\y+\sigma^-\y}{2}\nonumber\\
&=&c_0+c_1\ln\y+c_2\ln^2\y+\beta_{\cal P'}\y^{\mu-1},\\
\Delta\sigma&=&\frac{\sigma^{+}\y-\sigma^-\y}{2}\nonumber\\
&=&\delta\y^{\alpha -1},\\
m_{\rm av}&=&\frac{1}{2}\left(\frac{d\sigma^{+}}{d\x}+\frac{d\sigma^{-}}{d\x}\right)_{\nu =\nu_0}\nonumber\\
&=&c_1\left\{\frac{1}{\y}\right\}+c_2\left\{ \frac{2\ln\y}{\y}\right\}+\beta_{\cal P'}\left\{(\mu-1)\y^{\mu-2}\right\},\\
\Delta m&=&\frac{1}{2}\left(\frac{d\sigma^{+}}{d\x}-\frac{d\sigma^{-}}{d\x}\right)_{\nu =\nu_0}\nonumber\\
&=&\delta\left\{(\alpha -1)\y^{\alpha - 2}\right\}.
\end{eqnarray}
Using the definitions of $\sigma_{\rm av}$, $\Delta\sigma$, $m_{\rm av}$ and $\Delta m$, we now write the four constraint equations
\begin{eqnarray}
\beta_{\cal P'}&=&\frac{\y^{2-\mu}}{\mu -1}\left[m_{\rm av}-c_1\left\{\frac{1}{\y}\right\} -c_2\left\{\frac{2\ln\y}{\y}
\right\}\right],\label{deriveven}\\
c_0&=& \sigma_{\rm av}-c_1\ln\y-c_2\ln^2\y-\beta_{\cal P'}\y^{\mu-1},\label{intercepteven}\\
\alpha&=&1+\frac{\Delta m}{\Delta \sigma}\y,\label{derivodd}\\
\delta&=&\Delta \sigma\y^{1-\alpha}\label{interceptodd},
\end{eqnarray}
that utilize the two slopes and the two intercepts at the transition energy $\nu_0$, where we join on to the asymptotic fit. We pick $\nu_0$ as the (very low) energy just after which resonance behavior finishes. We use $\mu=0.5$ throughout, which is appropriate for a Regge-descending trajectory.  In the above, $m=m_p$ is the proton mass.

Our strategy is to use the rich amount of low energy data to constrain our  high energy fit. At the transition energy $\nu_0=7.59$ GeV, corresponding to  a cms (center of mass)  energy of $\sqrt s_0=4$ GeV,  the cross sections $\sigma^+(\nu_0/m)$ and $\sigma^-(\nu_0/m)$, along with the slopes $\left(\frac{d\sigma^{+}}{d(\x}\right)_{\nu=\nu_0}$ and $\left(\frac{d\sigma^{-}}{d(\x}\right)_{\nu=\nu_0}$, are used to constrain the asymptotic high energy fit so that it matches the low energy data at $\nu_0$. We picked $\nu_0$  much below the energy at which we start our high energy fit, but at an energy safely above the resonance regions. Very local fits are made to the region about the energy $\nu_0$ in order to evaluate the two cross sections and their two derivatives at $\nu_0$ that are needed in the above constraint equations. We next impose the 4 constraint equations, Equations (\ref{deriveven}), (\ref{intercepteven}), (\ref{derivodd}) and (\ref{interceptodd}), which we use in  our $\chi^2$ fit to Equations {\ref{sigmapm} and \ref{rhopm}. For safety, we  start the  data fitting at an energy $\nu_{\rm min}=18.25$ GeV, corresponding to the cms energy, $\sqrt s_{\rm min}=6.0$ GeV, appreciably higher than the transition energy.  
 
 We stress that the odd amplitude parameters $\alpha$ and $\delta$ and hence the odd amplitude itself, $\delta\x^{\alpha - 1}$,  are {\em completely determined} by the experimental values $\Delta m$ and $\Delta \sigma$ at the transition energy $\nu_0$. Thus, at {\em all} energies, the {\em differences} of the cross sections $\sigma^- -\sigma^+$ (from the optical theorem, the differences in the imaginary portion of the scattering amplitude) and the {\em differences} of the real portion of the scattering amplitude are completely fixed {\em before} we make our fit.  Further, for a $\ln^2s$ $(\ln s$) fit, the even amplitude parameters $c_0$ and $\beta_{\cal P}'$ are determined by $c_1$ and $c_2$ ($c_1$ only) along with the experimental values of $\sigma_{\rm av}$ and $m_{\rm av}$ at the transition energy $\nu_0$. In particular, for a $\ln^2s$ ($\ln s$) fit, we only fit the 3 (2) parameters $c_1$, $c_2$, and $f(0)$ ($c_1$ and $f_+(0))$.  Since the subtraction constant $f_+(0)$ only enters into the $\rho$-value determinations, only the 2  parameters  $c_1$ and $c_2$  of the original 7 are required for a $\ln^2s$  fit to the cross sections $\sigma^{\pm}$, which gives  us exceedingly little freedom in this fit---it is indeed very tightly constrained, with not much latitude for adjustment.  The cross sections  $\sigma^{\pm}$ for the $\ln s$ fit are even more tightly constrained, with only one adjustable parameter, $c_1$.

Table \ref{table:ppfitnew} summarizes the results of our simultaneous fits---using the 4 constraint equation---to the available accelerator data  from the Particle Data Group\cite{pdg} for  $\sigma_{pp}$, $\sigma_{\bar pp}$, $\rho_{pp}$ and $\rho_{\bar pp}$, after using the ``Sieve'' algorithm. Two $\delchimax$ cuts,  6 and 9, were made for $\ln^2(\nu/m)$ fits. The probability of the fit for the cut $\delchimax=6$ was $\sim 0.2$, a very satisfactory probability for this  many degrees of freedom, and we chose this data set. As seen in Table \ref{table:ppfitnew}, the fitted parameters are very insensitive to this choice.
\begin{table}[h,t]                   
%
\def\arraystretch{1.1}            
     \caption{\protect\small The fitted results for a 3-parameter $\chi^2$ fit with $\sigma\sim\ln^2(\nu/m_p)$ and a 2-parameter fit with $\sigma\sim\ln(\nu/m_p)$ to the total cross sections and $\rho$-values for $pp$ and $\bar pp$ scattering. The renormalized $\chi^2/\nu_{\rm min}$,  taking into account the effects of the $\delchimax$ cut, is given in the row  labeled ${\cal R}\times\chi^2_{\rm min}/\nu$. The errors in the fitted parameters have been multiplied by the appropriate $r_{\chi2}$.  The proton mass  is $m_p$ and the laboratory nucleon energy is $\nu$. \label{table:ppfitnew}}
\vspace {2mm}
\small
\begin{center}
\begin{tabular}[b]{|l||c|c||c||}
\cline{2-4}
\multicolumn{1}{c|}{}&\multicolumn{2}{c||}{$\sigma\sim \ln^2(\nu/m_p)$}&\multicolumn{1}{c||}{$\sigma\sim \ln(\nu/m_p)$}\\
\cline{1-1}
\multicolumn{1}{|c||}{Parameters }
      &\multicolumn{2}{|c||}{$\delchimax$}&\multicolumn{1}{|c||}{$\delchimax$}\\ 
\cline{2-4}
	\multicolumn{1}{|c||}{}
      &\multicolumn{1}{c|}{6}&\multicolumn{1}{c||}{9} &\multicolumn{1}{c||}{6}\\
      \hline
	\multicolumn{4}{|c||}{\ \ \ \ \  Even Amplitude}\\
	\cline{1-4}
      $c_0$\ \ \   (mb)&$37.32$ &$37.25$&28.26\\ 
      $c_1$\ \ \   (mb)&$-1.440\pm0.070$ &$-1.416\pm0.066$&$2.651\pm 0.0070$\\ 
	$c_2$\ \ \ \   (mb)&$0.2817\pm0.0064$&$0.2792\pm0.0059$&------\\
      $\beta_{\cal P'}$\ \   (mb)&$37.10$ &$37.17$&47.98\\ 
      $\mu$&$0.5$ &$0.5$&0.5\\ 
	$f(0)$ (mb GeV)&$-0.075\pm0.59$&$-0.069\pm 0.57$&$4.28\pm 0.59$\\
      \hline
	\multicolumn{4}{|c||}{\ \ \ \ \  Odd Amplitude}\\
	\hline
      $\delta$\ \ \   (mb)&$-28.56$ &$-28.56$&-28.56\\
      $\alpha$&$0.415$ &$0.415$&0.415\\ 
	\cline{1-4}
     	\hline
	\hline
	$\chi^2_{\rm min}$&181.6&216.6&2355.7\\
	${\cal R}\times\chi^2_{\rm min}$&201.5&222.5&2613.7\\ 
	$\nu$ (d.f).&184&189&185\\
\hline
	${\cal R}\times\chi^2_{\rm min}/\nu$&1.095&1.178&14.13\\
\hline
\end{tabular}
\end{center}
\end{table}
\def\arraystretch{1}  
 The same data set ($\delchimax=6$ cut) was also used for the $\ln(\nu/m)$ fit. The probability of the $\ln(\nu/m)$ fit is $<<10^{-16}$ and is clearly ruled out. This is illustrated graphically in \fig{fig:sigmapp}a.

We note  that when using a $\ln^2(\nu/m)$ fit {\em before} imposing the ``Sieve'' algorithm,  a value of $\chi^2$/d.f.=5.657 for 209 degrees of freedom was found, compared to $\chi^2$/d.f.=1.095 for 184 degrees of freedom when using the $\delchimax=6$ cut. The ``Sieve'' algorithm eliminated 25 points with energies $\sqrt s\ge6$ GeV (5 $\sigma_{pp}$, 5 $\sigma_{\pbar p}$, 15 $\rho_{pp}$), while changing the total renormalized $\chi^2$ from 1182.3 to 201.4. These 25 points that were screened out had a $\chi^2$ contribution of 980.9, an average value of 39.2. For a Gaussian distribution, about 3 points with $\delchi>6$ are expected, with a total $\chi^2$ contribution of slightly more than 18 and {\em not} 980.9. We see that  the ``Sieve'' algorithm has rid the two data sets of outliers.

Figure \ref{fig:sigmapp}a) shows the individual fitted cross sections (in mb) for $ pp$ and $\bar pp$ for 
$\ln^2(\nu/m)$ and $\ln(\nu/m)$ for the cut $\delchimax=6$ in Table \ref{table:ppfitnew}, plotted against the cms energy, $\sqrt s$, in GeV, and Figure \ref{fig:sigmapp}b) shows the individual fitted $\rho$-values for $pp$ and $\bar pp$  plotted against the cms energy, $\sqrt s$. The data shown are the sieved data with $\sqrt s \ge 6$ GeV. The $\ln^2(\nu/m)$ fits to the data sample with $\delchimax=6$, corresponding to the solid curve for $\bar pp$ and the dash-dotted curve for $pp$,  are excellent, yielding a total renormalized $\chi^2=201.5$, for 184 degrees of freedom, corresponding to a fit probability of $\sim0.2$. On the other hand, the $\ln(\nu/m)$ fits to the same data sample---the long dashed curve for $\bar p p$ and the short dashed curve for $pp$---are very bad fits, yielding a total $\chi^2=2613.7$ for 185 degrees of freedom, corresponding to a fit probability of $<<10^{-16}$. In essence, the $\ln(\nu/m)$ fit clearly undershoots {\em all} of the high energy cross sections. The ability of nucleon-nucleon scattering to distinguish cleanly  between an  energy dependence of $\ln^2(\nu/m_p)$ and  an energy dependence of $\ln(\nu/m_p)$ is quite dramatic.

A few remarks on our $\ln^2(\nu/m)$ asymptotic energy analysis for $pp$ and $\bar pp$ are in order. It should be stressed that we used {\em both} the CDF and E710/E811 high energy experimental cross sections at $\sqrt s=1800$ GeV in the $\ln^2(\nu/m)$ analysis. Inspection of Fig. \ref{fig:sigmapp}a) shows that at $\sqrt s=1800$ GeV, our fit  effectively passes below the  cross section point of $\sim$ 80 mb (CDF collaboration).  In particular, to test the sensitivity of our fit to the differences between the highest energy accelerator $\bar pp$ cross sections from the Tevatron, we next {\em omitted completely} the CDF ($\sim$ 80 mb) point and refitted the data without it.  This fit, also using $\delchimax=6$,  had a renormalized $\chi^2$/d.f.=1.055, compared to 1.095 with the CDF point included.  Since you only expect, on average, a $\Delta\chi^2$ of $\sim 1$ for the removal of one point, the removal of the CDF point slightly improved  the goodness-of-fit. Moreover, the new parameters of the fit were only {\em very minimally} changed. As an example, the predicted value from  the new fit for the cross section at $\sqrt s=1800$ GeV---{\em without} the CDF point---was $\sigma_{\bar pp}=75.1\pm0.6$ mb, where the error is the statistical error due to the errors in the  fitted parameters.   Conversely, the predicted value from 
Table \ref{table:predictions}---which used {\em both} the CDF and the E710/E811 point---was $\sigma_{\bar pp}=75.2\pm0.6$ mb, virtually identical. Further, at $\sqrt s=14$ TeV (LHC energy),  the fit {\em without} the  CDF point had $\sigma_{\bar pp}=107.2\pm1.2$, whereas {\em including} the CDF point (Table \ref{table:predictions}) gave $\sigma_{\bar pp}=107.3\pm1.2$. 
Thus, within errors, there was practically {\em no effect of either including or excluding} the CDF point. The fit was determined almost exclusively by  the E710/E811 cross section---presumably because the asymptotic fit was locked into the low energy transition energy $\nu_0$, thus sampling the rich amount of lower energy data.

In Table \ref{table:predictions}, we make high energy predictions of  total cross sections and $\rho$-values for $\bar pp$ and $pp$ scattering---from collider energies up to  the high energy regions appropriate to cosmic ray air shower experiments.
\begin{table}[h,t]                   
%

\bc
     \caption{\protect\small Predictions of high energy $\bar pp$ and $pp$ total  cross sections and $\rho$-values,  from Table \ref{table:ppfitnew}, $\sigma\sim\ln^2(\nu/m_\pi)$, $\delchimax=6$.\label{table:predictions}
}
\vspace{2mm}
\small

\begin{tabular}[h]{|l||c|c||c|c||}
    \cline{1-5}
      \multicolumn{1}{|l||}{ $\sqrt s$, in GeV}
      &\multicolumn{1}{c|}{$\sigma_{\bar pp}$, in mb}
      &\multicolumn{1}{c||}{$\rho_{\bar p p}$}&\multicolumn{1}{c|}{$\sigma_{ pp}$, in mb}&\multicolumn{1}{c||}{$\rho_{pp}$}\\

      \hline\hline
	540&$60.81\pm0.29$&$0.137\pm0.002$&$60.76\pm0.29$&$0.136\pm0.002$\\\hline
 	1,800&$75.19\pm0.55$&$0.139\pm0.001$&$75.18\pm0.55$&$0.139\pm0.001$\\\hline    
 	14,000&$107.3\pm1.2$&$0.132\pm0.001$&$107.3\pm1.2$&$0.132\pm0.001$\\\hline    
 	50,000&$132.1\pm1.7$&$0.124\pm0.001$&$132.1\pm1.7$&$0.124\pm0.001$\\\hline
 	100,000&$147.1\pm2.0$&$0.120\pm0.001$&$147.1\pm2.0$&$0.120\pm0.001$\\\hline
\end{tabular}
\ec
\end{table}

\def\arraystretch{1}  

We have demonstrated that the duality requirement that high energy cross sections smoothly interpolate into the resonance region strongly favors a $\ln^2s$ behavior of the asymptotic cross sections for the nucleon-nucleon systems, in agreement with  earlier result for $\gamma p$ scattering\cite{BH} and  $\pi p$ scattering\cite{igi, newFroissart}.  We conclude that the three hadronic systems, $\gamma p$, $\pi p$ and  nucleon-nucleon, {\em all} have an asymptotic $\ln^2s$ behavior, thus saturating the Froissart bound.

At 14 TeV, we predict  $\sigma_{\bar pp}=107.3\pm1.1$ mb and $\rho_{\bar pp}=0.132\pm 0.001$ for the Large Hadron Collider---robust predictions that rely critically  on the saturation of the Froissart bound.

Figure \ref{fig:sigppallenergies} shows all available data for both $\bar p p$ and $pp$, including cosmic ray data previously analyzed by Block, Halzen and Stanev\cite{BHS}. It is most striking that the two fitted curves for $\sigma_{nn}$even, using on the one hand, the $\ln^2(\nu/m)$ model of this work and on the other hand, the QCD-inspired model of the BHS group\cite{BHS}, are virtually indistinguishable over 5 decades of cms energy, {\em i.e.,} in the energy region $3\le \sqrt s\le 10^5$ GeV.

\section{Part 3: The Glauber analysis for $\sigma_{\rm p-air}$}
The extraction of the pp cross section from the cosmic ray data is a two
 step process. First, one calculates the $p$-air total cross section from
 the measured production cross section
\begin{equation}
\sigma_{\rm p-air}^{\rm prod} = \spa - \spae - \spaqe \,.  \label{eq:spa}
\end{equation}
 Next, the Glauber method\cite{glauber,gaisser} is used to transform the measured
 value of $\spai$ into a proton--proton total cross section $\sigma_{pp}$;
 all the necessary steps are calculable in the theory. In Eq.\,(\ref{eq:spa})
 the measured cross section for particle production is supplemented with
$\spae$ and $\spaqe$, 
 the elastic and quasi-elastic cross section, respectively, as calculated by the
 Glauber theory, to obtain the total cross section $\spa$. The subsequent
 relation between $\spai$ and $\sigma_{pp}$ involves $B$, the slope of the
 forward scattering amplitude for elastic $pp$ scattering, ${d\sigma_{pp}^{\rm el}\over dt}$, where 
$
B \equiv \left[ {d\over dt} \left(\ln{d\sigma_{pp}^{\rm el}\over dt}\right)
 \right]_{t=0} \,,
$
 and is shown in Fig.\,\ref{fig:B&sigpp}, which plots $B$ against
 $\sigma_{pp}$, for 5 curves of different values of $\spai$. 
 This summarizes the reduction procedure
 from $\spai$ to $\sigma_{pp}$~\cite{engel}. 

The solid curve used the value of $B$  found from the QCD-inspired fit of Block, Halzen and Stanev\cite{BHS}, whereas the $pp$ cross section value was found from the $\chi^2$ fit of Table \ref{table:ppfitnew}, for $\sigma\sim\ln^2(\nu/m)$ and $\delchimax=6$. The open circle is our value for $\sigma_{pp}$ at $\sqrt s=72.0$ TeV, the HiRes energy\cite{HiRes}. 

Unlike the Glauber calculation that was used in references \cite{engel} and \cite{BHS}, we have here incorporated inelastic screening into our calculation of \eq{eq:spa}, using a two-channel model to approximate diffraction. The screening  has the effect of lowering the observed p-air cross section, $\sigma_{\rm p-air}^{\rm prod}$, by about 30 mb at $\sqrt s=70$ GeV. In Fig. \ref{fig:pair&energy} the solid line is a plot of $\sigma_{\rm p-air}^{\rm prod}$ calculated including  inelastic screening, against the cms energy $\sqrt s$. The HiRes point at 72.0 TeV, the open diamond, is in good agreement with our prediction of $\sigma_{\rm p-air}^{\rm prod}$.

The Fly's Eye\cite{fly} and AGASA\cite{akeno}
 cosmic ray experiments  measure the
 shower attenuation length ($\Lambda_m$) and not the interaction length of the protons in the atmosphere
 ($\lpa$).  They {\em calculated} $\lpa$, and thus, $\sigma_{\rm p-air}^{\rm prod}$,  from the relation
\begin{equation}
\Lambda_m = k \lpa = k { 14.5 m \over {\sigma_{\rm p-air}^{\rm prod}}} \,,  \label{eq:Lambda_m}
\end{equation}
 where $k$  depends on the rate at which the energy of the primary proton
 is dissipated into electromagnetic shower energy observed in the
 experiment. The latter effect is parameterized in Eq.\,(\ref{eq:Lambda_m})
 by the parameter $k$; $m$ is the proton mass and $\sigma_{\rm p-air}^{\rm prod}$ the inelastic
 proton-air cross section. The values of $k$ in the original publications of ref. \cite{fly,akeno} were determined by Monte Carlo simulations which did not take into account scaling violations---Fly's Eye used $k=1.6$ and AGASA used $k=1.5$.  More modern Monte Carlo's such as the SIBYLL simulation\cite{sibyll} would give $k=1.2$.  We have {\em renormalized} the $k$ values to fit our curve in Fig. \ref{fig:pair&energy}, using $k=1.27$, in agreement with modern simulations. 

The HiRes point uses a different analysis method\cite{HiRes} which does {\em not} require a $k$ factor, being a more absolute measurement of $\sigma_{\rm p-air}^{\rm prod}$. Thus, the agreement of their value of $\sigma_{\rm p-air}^{\rm prod}$ with our prediction is of much more significance---the Fly's Eye and AGASA results are only shown because they can easily be made compatible with the predictions of Fig. \ref{fig:pair&energy} by rescaling $k$ to a more modern value, {\em i.e.,} $k=1.27$.

Finally, we show in Fig. \ref{fig:pp&&pair}
 the dependence of the total $pp$ cross section, $\sigma_{pp}$ on the observed p-air production cross section, $\sigma_{\rm p-air}^{\rm prod}$.

\section{Conclusions}Using the saturation of the Froissart bound for $pp$ scattering gives us a robust prediction of the observed p-air production cross section, $\sigma_{\rm p-air}^{\rm prod}$, when we use the $pp$ extrapolations to cosmic ray energies in a Glauber calculation with inelastic screening---thus tying together accelerator measurements with precise energy scales to cosmic ray measurements where the energy scale has rather large uncertainties.  The available cosmic ray measurements   of  $\sigma_{\rm p-air}^{\rm prod}$ are in reasonable agreement with our predictions.  Clearly, more accurate measurements of $\sigma_{\rm p-air}^{\rm prod}$ are needed at several different  energies, and we urge cosmic ray experimenters to make new efforts in these directions.

\bigskip
\section*{Acknowledgments} {\small We thank the Aspen Center for Physics, Aspen, Colorado,  for its hospitality while writing this paper.}
\bigskip


\begin{figure} 
\begin{center}
\mbox{\epsfig{file=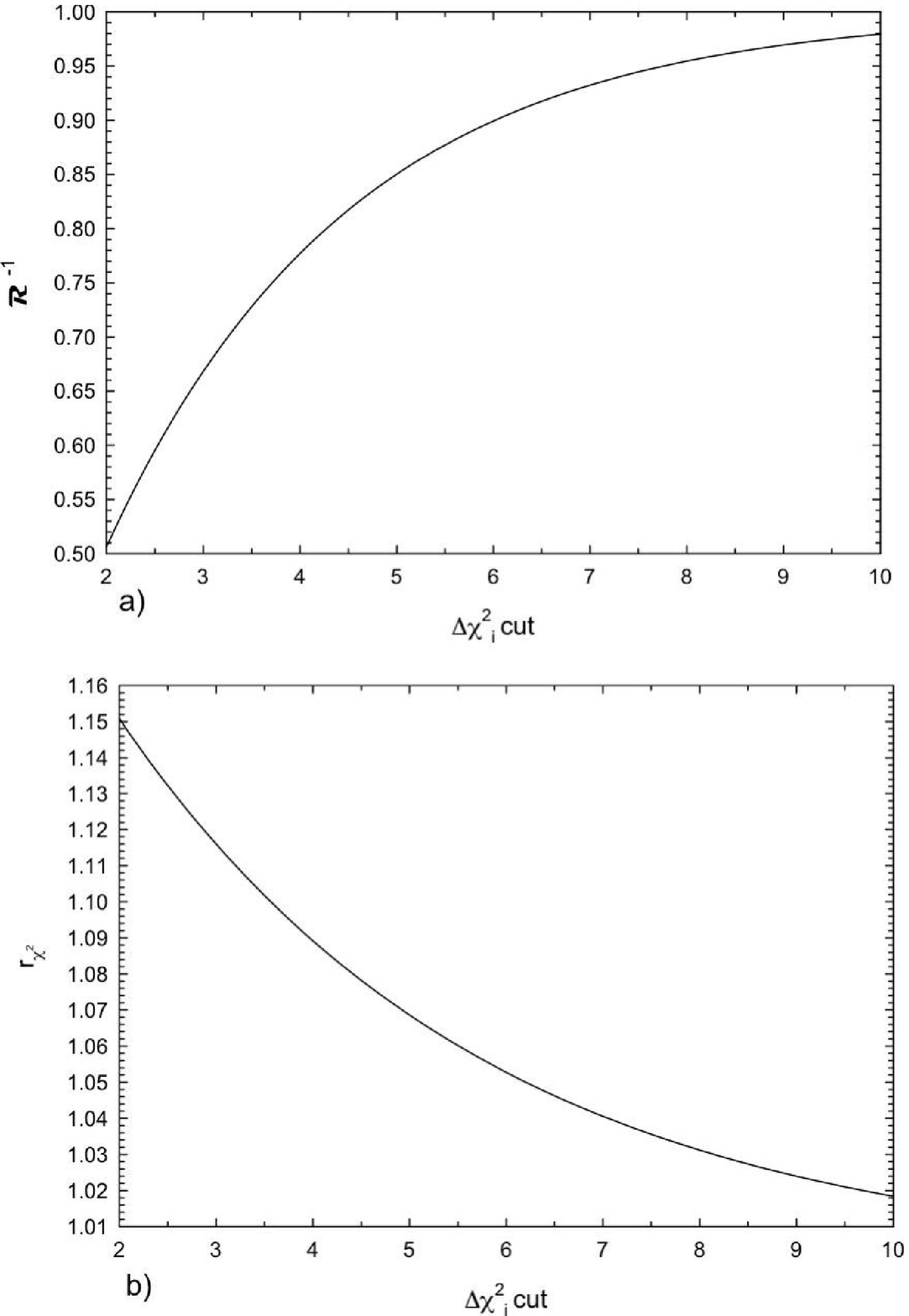,width=5in%
,bbllx=0pt,bblly=0pt,bburx=420pt,bbury=610pt,clip=%
}}
\end{center}
\caption[]{ \footnotesize 
a) A plot of  ${\cal R}^{-1}$, the reciprocal of the factor  that multiplies $\chi^2_{\rm min}/\nu$ found in the $\chi^2$ fit to the sifted data set  {\em vs.} $\delchi$ cut, {\em i.e.,} $\delchimax$.
\mbox{\ \ } b) A plot of $r_{\chi^2}$, the  factor whose square multiplies the covariant matrix found in the $\chi^2$ fit to the sifted data set  {\em vs.} $\delchi$ cut, {\em i.e.,} $\delchimax$. \noindent These figures are taken from ref. \cite{sieve}.  
  }
\label{renorm}
\end{figure}

\begin{figure} 
\begin{center}
\mbox{\epsfig{file=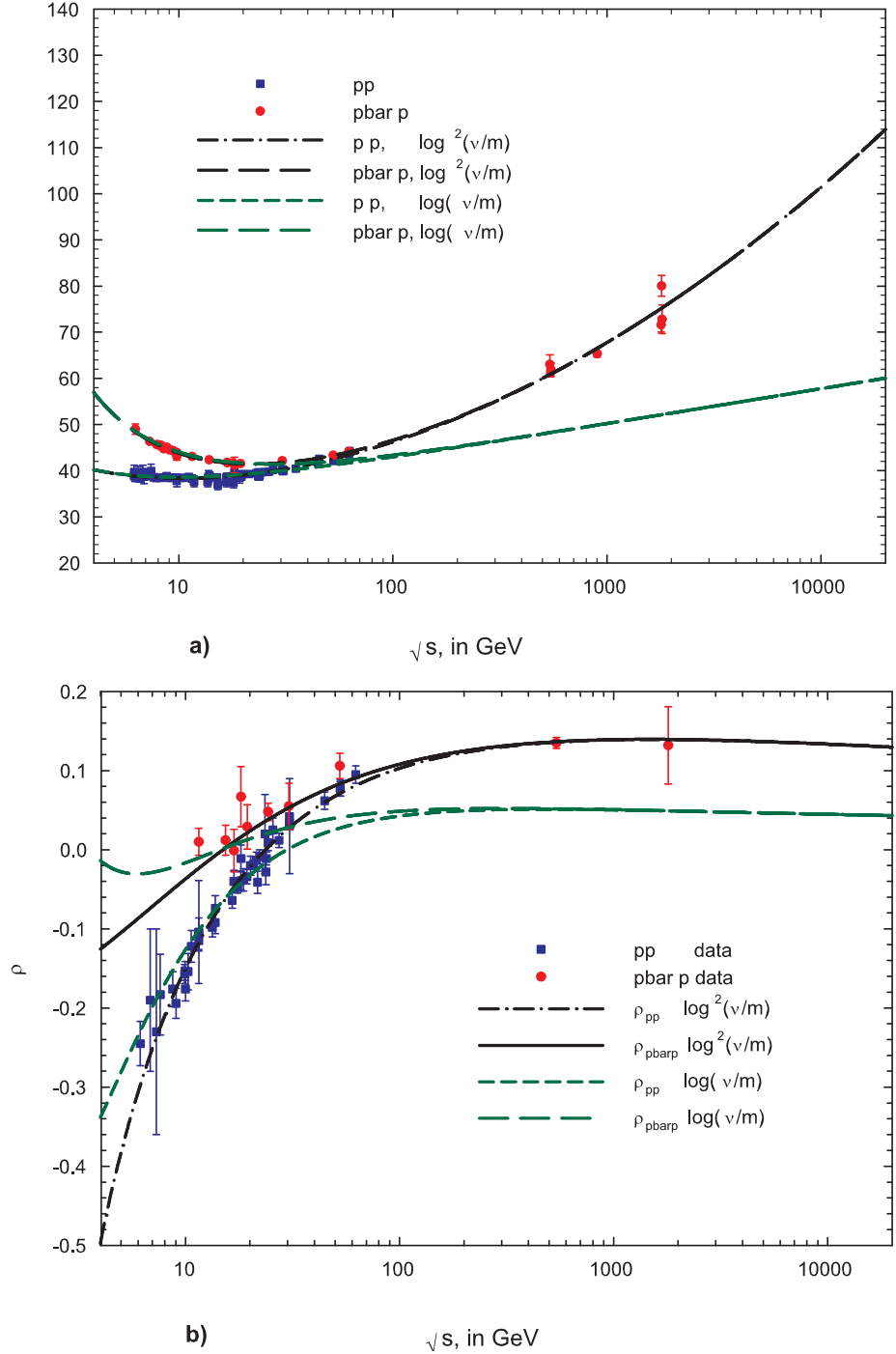,width=5in%
,bbllx=135pt,bblly=32pt,bburx=556pt,bbury=649pt,clip=%
}}
\end{center}
\caption[]{ \footnotesize
\ref{fig:sigmapp}a): The fitted total cross sections $\sigma_{p p}$ and $\sigma_{\pbar p}$ in mb, {\em vs.} $\sqrt s$, in GeV;\ \ref{fig:sigmapp}b): The fitted $\rho$-values,   $\rho_{p p}$ and $\rho_{\pbar p}$ {\em vs.} $\sqrt s$, in GeV,  using the 4 constraints of Equations (\ref{deriveven}), (\ref{intercepteven}), (\ref{derivodd}) and (\ref{interceptodd}).  The circles are the sieved data  for $\pbar p$ scattering and the squares are the sieved data for $p p$ scattering for $\sqrt s\ge 6$ GeV. The dash-dotted curve ($pp$)  and the solid curve ($\pbar p$) are the $\chi^2$ fits of Table \ref{table:ppfitnew}, for $\sigma\sim\ln^2(\nu/m)$ and $\delchimax=6$. The upper sign is for $p p$ and the lower sign is for $\pbar p$ scattering.  The short dashed curve ($p p$) and the long dashed curve ($\pbar p$) are $\chi^2$ fits of Table  \ref{table:ppfitnew} for $\sigma\sim\ln(\nu/m)$ and $\delchimax=6$. The laboratory energy of the nucleon is  $\nu$ and $m$ is the nucleon mass. 
  }
\label{fig:sigmapp}
\end{figure}
\begin{figure} 
\begin{center}
\mbox{\epsfig{file=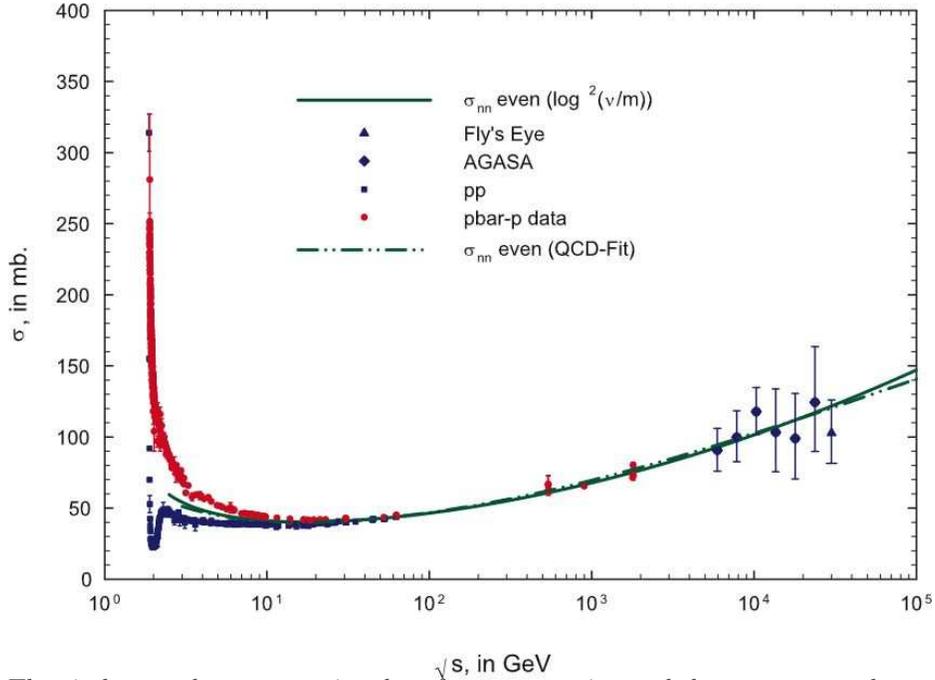,width=5.0in%
,bbllx=0pt,bblly=0pt,bburx=425pt,bbury=335pt,clip=%
}}
\end{center}
\vspace{-10mm}
\caption[]{ \footnotesize
The circles are the cross section data  for $\pbar p$ scattering and the squares are the cross section data for $p p$ scattering, in mb, {\em vs.} $\sqrt s$, in GeV, for all of the known accelerator data. The solid  curve is the $\chi^2$ fit (Table \ref{table:ppfitnew}, $\sigma\sim\ln^2(\nu/m)$, $\delchimax=6$) of  the high energy data  of the crossing-even amplitude, of the form~: $\sigma_{nn}{\rm even}=c_0 +c_1{\ln }\left(\nu\over m \right)+c_2{\ln }^2\left(\nu\over m)\right)+\beta_{\cal P'}\left(\nu\over m \right)^{\mu -1}$, with  $c_0$ and $\beta_{\cal P'}$ constrained by \eq{deriveven} and \eq{intercepteven}.    The dot-dot-dashed curve is the crossing-even amplitude cross section $\sigma_{nn}$, from a QCD-inspired fit that fit not only  the accelerator $\bar pp$ and $pp$ cross sections and $\rho$-values, but also fit the AGASA and Fly's Eye cosmic ray pp cross sections.

}
\label{fig:sigppallenergies}
\end{figure}
\begin{figure} 
\begin{center}
\mbox{\epsfig{file=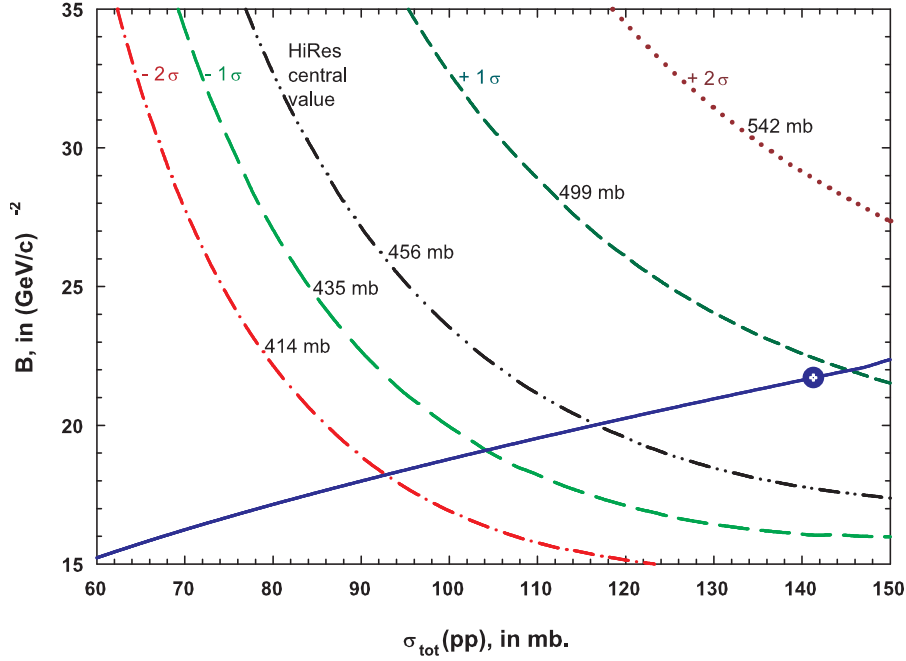,width=5.0in%
,bbllx=90pt,bblly=241pt,bburx=525pt,bbury=580pt,clip=%
}}
\end{center}
\vspace{-.2in}
\caption[]{ \footnotesize
$B$ dependence on the total cross section, $\sigma_{\rm tot}(pp)$.  The five curves are lines of constant $\sigma_{\rm p-air}^{\rm prod}$, of 414, 435, 456, 499 and 542 mb---the central value is the HiRes\cite{HiRes} value and the others are $\pm 1\sigma$ and $\pm 2\sigma$.  The solid curve is the QCD-inspired\cite{BHS} fit of $B$ against the $\ln^2 s$ fit of $\sigma_{\rm tot}(pp)$ from Table \ref{table:ppfitnew} for $\delchimax=6$.  The open circle is our value for $\sqrt s=72.0$ TeV, the HiRes energy.
}
\label{fig:B&sigpp}
\end{figure}
%
\begin{figure} 
\begin{center}
\mbox{\epsfig{file=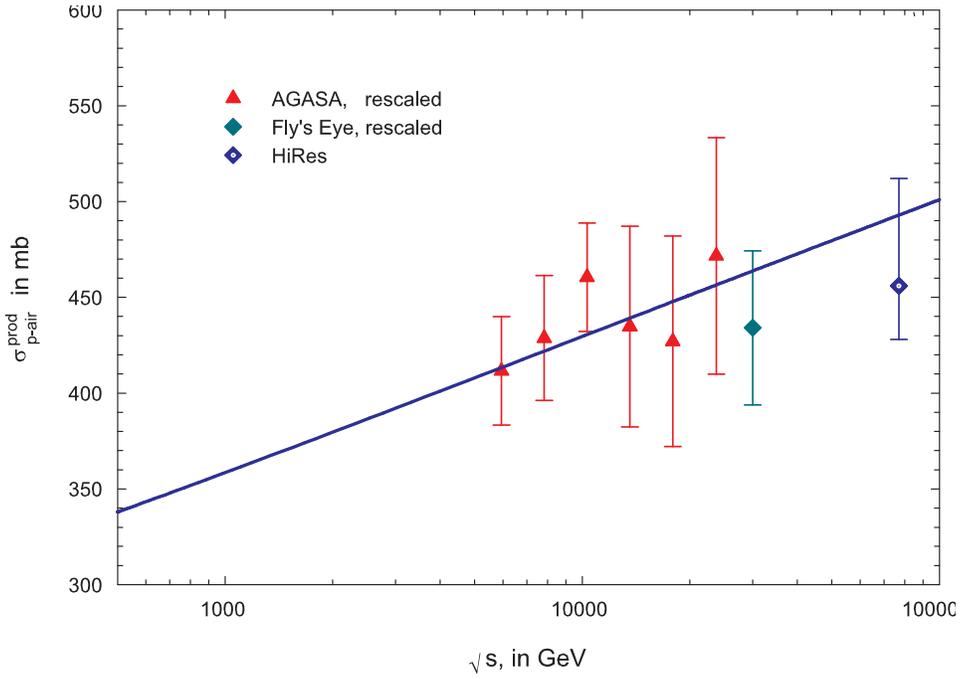,width=5.0in%
,bbllx=0pt,bblly=50pt,bburx=420pt,bbury=345pt,clip=%
}}
\end{center}
\vspace{-.2in}
\caption[]{ \footnotesize
The p-air production cross section, $\sigma_{\rm p-air}^{\rm prod}$, in mb.
{\em vs.} the cms energy, $\sqrt s$, in GeV.
}
\label{fig:pair&energy}
\end{figure}
%

 
\begin{figure}[h] 
\begin{center}

\mbox{\epsfig{file=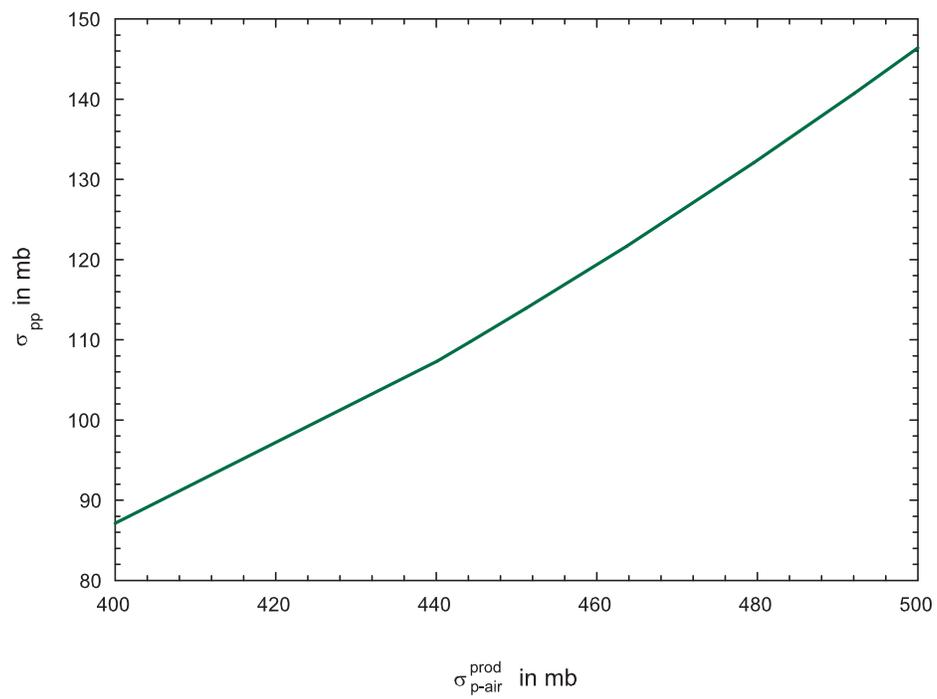,width=5.0in%
,bbllx=0pt,bblly=0pt,bburx=430pt,bbury=310pt,clip=%
}}
\end{center}
\vspace{-.2in}
\caption[]{ \footnotesize
The total $pp$ cross section, $\sigma_{pp}$, in mb {\em vs.} the p-air production cross section, $\sigma_{\rm p-air}^{\rm prod}$, in mb.
}
\label{fig:pp&&pair}
\end{figure}
%


%
%
%
\end{document}